# Effect of magnetic field on viscosity in the weakly ionized and magnetized plasma with power-law q-distributions in nonextensive statiostics

Yue Wang and Jiulin Du

*Department of Physics, School of Science, Tianjin University, Tianjin 300072, China*

**Abstract** We study the effect of magnetic field on the viscosity of charged particles in the weakly ionized and magnetized plasma with the power-law $q$-distributions by using the generalized Boltzmann equation of transport in nonextensive statistics. We derive six viscosity coefficients of electrons and ions in the complex plasma, including the magnetic field introduced four viscosity coefficients, where the $q$-parameter and magnetic field both play significant roles. By numerical analyses, we show that these viscosity coefficients depend strongly on the magnetic field and $q$-parameters of the weakly ionized and magnetized plasma, and so have potential applications in low temperature complex plasmas.

## 1. Introduction

Power-law distributions are a class of non-Maxwellian distributions commonly existed in the wide fields of nonequilibium complex systems in physics, astronomy, chemistry, biology and engineering technology etc. One example is the $q$-distributions in complex systems described within the framework of nonextensive statistics [1],

$$f_q(\mathbf{v}) = nB_q \left(\frac{m}{2\pi k_B T}\right)^{\frac{3}{2}} \left[1-(1-q)\frac{m(\mathbf{v}-\mathbf{u})^2}{2k_B T}\right]^{1/(1-q)}, \qquad (1)$$

where **v** is the velocity, $q$ is the nonextensive parameter who's deviation from unity represents the degree of nonextensivity, $T$ is temperature, **u** is the bulk velocity of the fluid under consideration, $n$ is the number density of particles, $m$ is mass of particle, $k_B$ is Boltzmann constant, and $B_q$ is a $q$-dependent normalized constant,

$$B_q = \begin{cases} (1-q)^{\frac{1}{2}}(3-q)(5-3q)\dfrac{\Gamma\left(\frac{1}{2}+\frac{1}{1-q}\right)}{4\Gamma\left(\frac{1}{1-q}\right)}, & \text{for } q \leq 1. \\[2ex] (q-1)^{\frac{3}{2}}\dfrac{\Gamma\left(\frac{1}{q-1}\right)}{\Gamma\left(\frac{1}{q-1}-\frac{3}{2}\right)}, & \text{for } q \geq 1. \end{cases}$$

If we let $q\rightarrow 1$ the velocity $q$-distribution (1) can become a Maxwell distribution.

Another example of power-law distributions is the $\kappa$-distributions often observed and studied in astrophysical and space plasmas [2-6], such as

$$f_\kappa(\mathbf{v}) = nB_\kappa \left(1+(2\kappa-3)^{-1}\frac{mv^2}{k_B T}\right)^{-(\kappa+1)}, \qquad (2)$$

where the parameter $3/2 < \kappa < \infty$ describes the distance deviated from the Maxwellian distribution, $B_\kappa$ is the $\kappa$-dependent normalization constant,

$$B_\kappa = \left[\frac{2\pi k_B T}{m}(\kappa - \frac{3}{2})\right]^{-3/2} \frac{\Gamma(\kappa+1)}{\Gamma(\kappa - \frac{1}{2})}.$$

Obviously, in the limit of the parameter $\kappa \to \infty$, the $\kappa$-distribution (2) reduces to a Maxwellian distribution.

One more example of power-law distributions is the $\delta$-distributions existed in various fields of natural science including plasmas [7-12], such as

$$f_\delta(E) \sim E^{-\delta}, \tag{3}$$

with the parameter $\delta > 0$, and the energy $E$. In theory, it is shown [13] that such a $\delta$-distribution is actually a special case of the $q$-distributions and/or the $\kappa$-distributions when there exists a fluctuation-dissipation relation such as $D = \delta^{-1} m\gamma E$ between the diffusion coefficient $D$ and the friction coefficient $\gamma$ in a complex dynamical system.

All the power-law distributions have found their dynamical origins and conditions from the Fokker-Planck equations for complex systems [13]. And the power-law distributions such as Eqs.(1)-(3) in complex systems can be described under the framework of nonextensive statistical mechanics [1], for example, plasmas [14-42] and self-gravitating systems [15]. In general, $T$, $n$ and $\mathbf{u}$ in nonequilibrium complex plasma should be considered to be space inhomogeneous, i.e., $T=T(\mathbf{r})$, $n=n(\mathbf{r})$ and $\mathbf{u}=\mathbf{u}(\mathbf{r})$, and they can vary with time. The key parameters, for example, the $\kappa$- and/or $q$-parameter for the electromagnetic interactions of the $q$-distributed nonequlibrium plasma is given by the equation [14,16]:

$$k_B \nabla T = e(1-q)(\nabla\varphi - c^{-1}\mathbf{u}\times\mathbf{B}), \tag{4}$$

where $e$ is electron charge, $\varphi$ is the Coulomb potential, $\mathbf{B}$ is the magnetic induction and $c$ is the light speed. Therefore, the $q$-distribution (1) for $q \neq 1$ can describe the properties of the magnetized complex plasma at a nonequiibrium stationary-state.

The weakly ionized plasma is of low temperature plasmas. In research fields of plasmas physics, the power-law distributions like $\kappa$- and $q$-distributions have been observed and studied widely in astrophysical plasmas, space plasmas and laboratory plasmas. For instance, in recent years, a variety of waves and instabilities [17-28] and many new properties of the power-law distributed complex plasmas [29-37] etc. have aroused great interest and wide attention. The transport coefficients in the $\kappa$-distributed plasmas were first studied under a very simplified Lorentz model [38]. Most recently, the diffusion in the weakly ionized plasma with the $\kappa$-distributions [39] and some transport properties of the full/weakly ionized plasma with the $\kappa$- and $q$-distributions were studied [40-42]. In this work, we study the viscosity coefficients in the weakly ionized and magnetized plasma with the power-law $q$-distributions, and we focus on the effect of magnetic field on the viscosity in the $q$-distributed complex plasma.

The paper is organized as follows. In Sec.2, we introduce the generalized Boltzmann equation of transport for the weakly ionized and magnetized complex plasma with the power-law $q$-distributions in nonextensive kinetics, and then derive the nonequilibrium $q$-distribution functions with the magnetic field. In Sec.3, we focus on the role of magnetic field in the viscosity of charged particles in the $q$-distributed plasma to derive the first and second viscosity coefficients with magnetic field. Finally in Sec.4, we give the conclusion.

## 2. The generalized transport equations and $q$-distribution functions

As we know, for the weakly ionized and magnetized plasma with the power-law $q$-distributions, when the magnetic field is taken into consideration, the generalized Boltzmann equation of transport in nonextensive kinetics can be written [42] by

$$\frac{\partial f_\alpha}{\partial t} + \mathbf{v} \cdot \frac{\partial f_\alpha}{\partial \mathbf{r}} + \frac{Q_\alpha}{m_\alpha}\left[\mathbf{E} + c^{-1}\mathbf{v} \times \mathbf{B}\right] \cdot \frac{\partial f_\alpha}{\partial \mathbf{v}} = C_q(f_\alpha), \tag{5}$$

where $f_\alpha \equiv f_\alpha(\mathbf{r}, \mathbf{v}, t)$ is a single-particle distribution function at time $t$, velocity $\mathbf{v}$, and position $\mathbf{r}$, the subscript $\alpha$ denotes electron and ion, $\alpha = e, i$, respectively. $Q_\alpha$ is electric charge of the $\alpha$th component, and $\mathbf{E}$ is the electric field intensity. In the following, we appoint that $\mathbf{r} = (r_x, r_y, r_z) \equiv (x, y, z)$. The term $C_q$ is the nonextensive $q$-collision term.

Following the same way of [42] for the viscosity in the plasma without magnetic field, if the $q$-collision term is considered as the generalized Krook model and in the first-order approximation of Chapman-Enskog expansion for the velocity distribution functions in the $q$-distributed plasma, $f_\alpha = f_{q,\alpha} + f_{q,\alpha}^{(1)}$, the transport equation (5) can be written as

$$\left\{\frac{\partial}{\partial t} + \mathbf{v} \cdot \frac{\partial}{\partial \mathbf{r}} + \frac{Q_\alpha}{m_\alpha}\left[\mathbf{E} + c^{-1}\mathbf{v} \times \mathbf{B}\right] \cdot \frac{\partial}{\partial \mathbf{v}}\right\}\left(f_{q,\alpha} + f_{q,\alpha}^{(1)}\right) = -\nu_\alpha f_{q,\alpha}^{(1)}, \tag{6}$$

where $\nu_\alpha$ is the mean collision frequency, which, as usual, can be considered constant, $f_{q,\alpha}^{(1)}$ is the first-order small disturbance about the stationary $q$-distribution $f_{q,\alpha}$, given for the $\alpha$th component by the power-law form [42],

$$f_{q,\alpha}(\mathbf{r},\mathbf{v}) = n_\alpha B_{q,\alpha} \left(\frac{m_\alpha}{2\pi k_B T_\alpha}\right)^{3/2} \left[1-(1-q_\alpha)\frac{m_\alpha}{2k_B T_\alpha}(\mathbf{v}-\mathbf{u})^2\right]^{1/(1-q_\alpha)}, \tag{7}$$

with

$$B_{q,\alpha} = \begin{cases} (1-q_\alpha)^{\frac{1}{2}}(3-q_\alpha)(5-3q_\alpha)\dfrac{\Gamma\left(\frac{1}{2}+\frac{1}{1-q_\alpha}\right)}{4\Gamma\left(\frac{1}{1-q_\alpha}\right)}, & \text{for } 0 < q_\alpha \leq 1. \\[1em] (q_\alpha-1)^{\frac{3}{2}}\dfrac{\Gamma\left(\frac{1}{q_\alpha-1}\right)}{\Gamma\left(\frac{1}{q_\alpha-1}-\frac{3}{2}\right)}, & \text{for } q_\alpha \geq 1. \end{cases}$$

Usually, the transport processes are studied in a steady state so that $\partial f_\alpha/\partial t = 0$. And if the first-order small disturbance satisfies $f_{q,\alpha}^{(1)} \ll f_{q,\alpha}$, we can neglect $f_{q,\alpha}^{(1)}$ on the left side of Eq.(6), and at the same time we retain the role of the magnetic field. Thus the transport equation (6) becomes

$$\mathbf{v} \cdot \frac{\partial f_{q,\alpha}}{\partial \mathbf{r}} + \frac{Q_\alpha \mathbf{E}}{m_\alpha} \cdot \frac{\partial f_{q,\alpha}}{\partial \mathbf{v}} = -\frac{Q_\alpha}{m_\alpha}c^{-1}\mathbf{v} \times \mathbf{B} \cdot \frac{\partial f_{q,\alpha}^{(1)}}{\partial \mathbf{v}} - \nu_\alpha f_{q,\alpha}^{(1)}, \tag{8}$$

where the equation $(\mathbf{v} \times \mathbf{B}) \cdot \partial f_{q,\alpha}/\partial \mathbf{v} = 0$ has been used because

$$\frac{\partial f_{q,\alpha}}{\partial \mathbf{v}} = -(\mathbf{v}-\mathbf{u})\frac{m_\alpha(1-q_\alpha)}{k_B T_\alpha}\left[1-(1-q_\alpha)\frac{m_\alpha}{2k_B T_\alpha}(\mathbf{v}-\mathbf{u})^2\right]^{-1} f_{q,\alpha}, \tag{9}$$

and without loss of generality, we can let $\mathbf{u}=0$ (but its derivatives do not equal zero)

[43-46] since the calculations for the viscosity coefficients in the matrix elements of stress tensor depend only on the derivatives of the velocity (see Eqs.(23)-(28)).

Substituting the $q$-distribution (7) into Eq.(8), we have that

$$\frac{m_\alpha}{k_B T_\alpha} \mathbf{v} \cdot \left( \mathbf{v} \cdot \frac{\partial \mathbf{u}}{\partial \mathbf{r}} \right) \left[ 1 - (1-q_\alpha) \frac{m_\alpha \mathbf{v}^2}{2k_B T_\alpha} \right]^{-1} f_{q,\alpha} = -\frac{Q_\alpha}{m_\alpha c} (\mathbf{v} \times \mathbf{B}) \cdot \frac{\partial f_{q,\alpha}^{(1)}}{\partial \mathbf{v}} - \nu_\alpha f_{q,\alpha}^{(1)}, \qquad (10)$$

where we have only retained the even function part of $\mathbf{v}$ because in calculations of the matrix elements of stress tensor, the integrations for odd function part of $\mathbf{v}$ equal zero (see Eq.(29)).

We let the direction of magnetic field be parallel to z-axis, $\hat{\mathbf{B}} = \mathbf{B}/B$, and $\omega_{Be} = Q_\alpha B/m_\alpha c$ be Larmor frequency, so that Eq.(10) is written as

$$\frac{m_\alpha}{k_B T_\alpha} \mathbf{v} \cdot \left( \mathbf{v} \cdot \frac{\partial \mathbf{u}}{\partial \mathbf{r}} \right) \left[ 1 - (1-q_\alpha) \frac{m_\alpha \mathbf{v}^2}{2k_B T_\alpha} \right]^{-1} f_{q,\alpha} = -\omega_{Be} \left( v_y \frac{\partial}{\partial v_x} - v_x \frac{\partial}{\partial v_y} \right) f_{q,\alpha}^{(1)} - \nu_\alpha f_{q,\alpha}^{(1)}. \qquad (11)$$

Now we can search the solution of Eq.(11) in the following form (only the even function part of $\mathbf{v}$) [43-45],

$$f_{q,\alpha}^{(1)} = k_{q1} v_x^2 + k_{q2} v_y^2 + k_{q3} v_z^2 + k_{q4} v_x v_y + k_{q5} v_x v_z + k_{q6} v_y v_z, \qquad (12)$$

where $k_{qi}$ ($i = 1, 2, \ldots, 6$) are undetermined functions. Substituting (12) into Eq.(11), we find that (see Appendix A),

$$k_{q1} = F_{q,\alpha} \frac{1}{\nu_\alpha} \left[ (\nu_\alpha^2 + 2\omega_{Be}^2) \frac{\partial u_x}{\partial x} + 2\omega_{Be}^2 \frac{\partial u_y}{\partial y} + \omega_{Be} \nu_\alpha \left( \frac{\partial u_x}{\partial y} + \frac{\partial u_y}{\partial x} \right) \right], \qquad (13)$$

$$k_{q2} = F_{q,\alpha} \frac{1}{\nu_\alpha} \left[ 2\omega_{Be}^2 \frac{\partial u_x}{\partial x} + (\nu_\alpha^2 + 2\omega_{Be}^2) \frac{\partial u_y}{\partial y} - \omega_{Be} \nu_\alpha \left( \frac{\partial u_x}{\partial y} + \frac{\partial u_y}{\partial x} \right) \right], \qquad (14)$$

$$k_{q3} = F_{q,\alpha} \frac{1}{\nu_\alpha} \frac{\partial u_z}{\partial z}, \qquad (15)$$

$$k_{q4} = F_{q,\alpha} \left[ \nu_\alpha \left( \frac{\partial u_x}{\partial y} + \frac{\partial u_y}{\partial x} \right) - 2\omega_{Be} \left( \frac{\partial u_x}{\partial x} - \frac{\partial u_y}{\partial y} \right) \right], \qquad (16)$$

$$k_{q5} = F_{q,\alpha} \left[ \nu_\alpha \left( \frac{\partial u_x}{\partial z} + \frac{\partial u_z}{\partial x} \right) + \omega_{Be} \left( \frac{\partial u_y}{\partial z} + \frac{\partial u_z}{\partial y} \right) \right], \qquad (17)$$

and

$$k_{q6} = F_{q,\alpha} \left[ \nu_\alpha \left( \frac{\partial u_y}{\partial z} + \frac{\partial u_z}{\partial y} \right) - \omega_{Be} \left( \frac{\partial u_x}{\partial z} + \frac{\partial u_z}{\partial x} \right) \right], \qquad (18)$$

where $u_x, u_y, u_z$ are the three rectangular axis components of the velocity $\mathbf{u} = (u_x, u_y, u_z)$, and we have denoted that

$$F_{q,\alpha} = -\frac{m_\alpha f_{q,\alpha}}{k_B T_\alpha (\nu_\alpha^2 + 4\omega_{Be}^2)} \left[ 1 - (1-q_\alpha) \frac{m_\alpha v^2}{2k_B T_\alpha} \right]^{-1}. \qquad (19)$$

Therefore, in the first-order approximation of Chapman-Enskog expansion, the velocity $q$-distribution functions for studying the viscosity of charged particles in the weakly ionized and magnetized plasma with the power-law $q$-distributions are expressed by the form,

$$f_\alpha = f_{q,\alpha} + f_{q,\alpha}^{(1)},$$
$$= f_{q,\alpha} + k_{q1}v_x^2 + k_{q2}v_y^2 + k_{q3}v_z^2 + k_{q4}v_xv_y + k_{q5}v_xv_z + k_{q6}v_yv_z. \tag{20}$$

## 3. The viscosity $q$-coefficients with the effect of magnetic field

According to the law of conservation of momentum, in the motion equation of plasma hydrodynamics, the stress tensor is $\mathbf{P}=\{P_{ij}\}_{i,j=x,y,z}$. When there is no magnetic field, the matrix elements can be given [42-44] by

$$P_{ij} = p_q \delta_{ij} - \eta_{q,0}\left(\frac{\partial u_i}{\partial r_j} + \frac{\partial u_j}{\partial r_i} - \frac{2}{3}\frac{\partial u_k}{\partial r_k}\delta_{ij}\right) - \zeta_q \frac{\partial u_k}{\partial r_k}\delta_{ij}, \tag{21}$$

where $\eta_{q,0}$ is the first viscosity coefficient, $\zeta_q$ is the second viscosity coefficient or volume viscosity coefficient, $p_q$ is the plasma static pressure, and according to the appointment we have that

$$\frac{\partial u_k}{\partial r_k} \equiv \sum_{k=x,y,z} \frac{\partial u_k}{\partial r_k}. \tag{22}$$

Eq. (21) combined with Eq.(22) is actually equal to a Navier-Stokes equation. In previous work, we derived $\eta_{q,0}$ and $\zeta_q$ in the $q$-distributed plasma [42].

However, in the magnetized plasma, the transport of momentum occurs at very different rates in different directions. In other words, the viscosity coefficients will depend on the direction of magnetic field. According to the previous designation that if the magnetic field $\mathbf{B}$ is parallel to z-axis, in this situation, the matrix elements of the stress tensor are changed as the following forms [43-46],

$$P_{xx} = p_q - \eta_{q,0}\left(\frac{\partial u_x}{\partial x} + \frac{\partial u_y}{\partial y} - \frac{2}{3}\frac{\partial u_k}{\partial r_k}\right) - \eta_{q,1}\left(\frac{\partial u_x}{\partial x} - \frac{\partial u_y}{\partial y}\right) - \eta_{q,3}\left(\frac{\partial u_x}{\partial y} + \frac{\partial u_y}{\partial x}\right) - \zeta_q\frac{\partial u_k}{\partial r_k}, \tag{23}$$

$$P_{yy} = p_q - \eta_{q,0}\left(\frac{\partial u_x}{\partial x} + \frac{\partial u_y}{\partial y} - \frac{2}{3}\frac{\partial u_k}{\partial r_k}\right) - \eta_{q,1}\left(\frac{\partial u_y}{\partial y} - \frac{\partial u_x}{\partial x}\right) + \eta_{q,3}\left(\frac{\partial u_x}{\partial y} + \frac{\partial u_y}{\partial x}\right) - \zeta_q\frac{\partial u_k}{\partial r_k}, \tag{24}$$

$$P_{xy} = P_{yx} = -\eta_{q,1}\left(\frac{\partial u_x}{\partial y} + \frac{\partial u_y}{\partial x}\right) + \eta_{q,3}\left(\frac{\partial u_x}{\partial x} - \frac{\partial u_y}{\partial y}\right), \tag{25}$$

$$P_{xz} = P_{zx} = -\eta_{q,2}\left(\frac{\partial u_x}{\partial z} + \frac{\partial u_z}{\partial x}\right) - \eta_{q,4}\left(\frac{\partial u_y}{\partial z} + \frac{\partial u_z}{\partial y}\right), \tag{26}$$

$$P_{yz} = P_{zy} = -\eta_{q,2}\left(\frac{\partial u_y}{\partial z} + \frac{\partial u_z}{\partial y}\right) + \eta_{q,4}\left(\frac{\partial u_x}{\partial z} + \frac{\partial u_z}{\partial x}\right), \tag{27}$$

$$P_{zz} = p_q - \eta_{q,0}\left(2\frac{\partial u_z}{\partial z} - \frac{2}{3}\frac{\partial u_k}{\partial r_k}\right) - \zeta_q\frac{\partial u_k}{\partial r_k}, \tag{28}$$

where due to the effect of magnetic field, the first viscosity coefficient becomes five parts: $\eta_{q,0}$, $\eta_{q,1}$, $\eta_{q,2}$, $\eta_{q,3}$, and $\eta_{q,4}$ are respectively the viscosity coefficients in different directions of the plasma affected by the magnetic field. As we expected, the element $P_{zz}$ parallel to the magnetic field is the same as Eq.(21), which has not been affected by the magnetic field.

Now, based on the integral expressions of the matrix element, defined [43-46] by

$$P_{\alpha,ij} = \int m_\alpha (v_i - u_i)(v_j - u_j) f_\alpha(\mathbf{r},\mathbf{v}) d\mathbf{v}, \tag{29}$$

we can calculate the effect of magnetic field on the viscosity coefficients of charged particles in the weakly ionized and magnetized plasma with the power-law velocity $q$-distributions.

Substituting (20) into (29) and letting $\mathbf{u}=0$ (but its derivatives do not equal zero) [43-46], we derive that (see Appendix B),

$$\begin{aligned} P_{\alpha,xx} &= m_\alpha \int d\mathbf{v} v_x^2 \left( f_{q,\alpha} + k_{q1}v_x^2 + k_{q2}v_y^2 + k_{q3}v_z^2 + k_{q4}v_xv_y + k_{q5}v_xv_z + k_{q6}v_yv_z \right) \\ &= \frac{2n_\alpha k_B T_\alpha}{(7-5q_\alpha)} - \frac{2n_\alpha k_B T_\alpha}{(7-5q_\alpha)(v_\alpha^2 + 4\omega_{Be}^2)v_\alpha} \left[ \left(3v_\alpha^2 + 8\omega_{Be}^2\right)\frac{\partial u_x}{\partial x} + \left(v_\alpha^2 + 8\omega_{Be}^2\right)\frac{\partial u_y}{\partial y} \right. \\ &\quad \left. + \left(v_\alpha^2 + 4\omega_{Be}^2\right)\frac{\partial u_z}{\partial z} + 2\omega_{Be}v_\alpha\left(\frac{\partial u_x}{\partial y} + \frac{\partial u_y}{\partial x}\right) \right], \quad 0 < q_\alpha < \frac{7}{5}. \end{aligned} \tag{30}$$

Since the static pressure of the nonextensive plasma gas [47] is

$$p_{q,\alpha} = \frac{2}{(7-5q_\alpha)} n_\alpha k_B T_\alpha, \quad 0 < q_\alpha < \frac{7}{5}, \tag{31}$$

comparing Eq. (30) with Eq. (23), we find the following viscosity coefficients for the $\alpha$th plasma component,

$$\eta_{q,0} = \frac{2}{(7-5q_\alpha)} \frac{n_\alpha k_B T_\alpha}{v_\alpha}, \tag{32}$$

$$\eta_{q,1} = \frac{2}{(7-5q_\alpha)} \frac{v_\alpha n_\alpha k_B T_\alpha}{(v_\alpha^2 + 4\omega_{Be}^2)}, \tag{33}$$

$$\eta_{q,3} = \frac{2}{(7-5q_\alpha)} \frac{2\omega_{Be} n_\alpha k_B T_\alpha}{(v_\alpha^2 + 4\omega_{Be}^2)}, \tag{34}$$

$$\zeta_q = \frac{2}{(7-5q_\alpha)} \frac{5 n_\alpha k_B T_\alpha}{3v_\alpha}. \tag{35}$$

In the nonextensive kinetic theory [47,48], limitations to value of the $q$-parameter exist generally. The above calculations and results hold true for $0 < q_\alpha < 7/5$, but for $q_\alpha \geq 7/5$, the calculations are diverges.

In the same way as the calculations for Eq. (30), we can derive that

$$\begin{aligned} P_{\alpha,xz} &= m_\alpha \int d\mathbf{v} v_x v_z \left( f_{q,\alpha} + k_{q1}v_x^2 + k_{q2}v_y^2 + k_{q3}v_z^2 + k_{q4}v_xv_y + k_{q5}v_xv_z + k_{q6}v_yv_z \right) \\ &= -\frac{2n_\alpha k_B T_\alpha}{(7-5q_\alpha)(v_\alpha^2 + \omega_{Be}^2)} \left[ v_\alpha\left(\frac{\partial u_x}{\partial z} + \frac{\partial u_z}{\partial x}\right) + \omega_{Be}\left(\frac{\partial u_y}{\partial z} + \frac{\partial u_z}{\partial y}\right) \right], \quad 0 < q_\alpha < \frac{7}{5}. \end{aligned} \tag{36}$$

Comparing Eq. (36) with Eq. (26), we find the viscosity coefficients for the $\alpha$th plasma component,

$$\eta_{q,2} = \frac{2}{(7-5q_\alpha)} \frac{v_\alpha n_\alpha k_B T_\alpha}{(v_\alpha^2 + \omega_{Be}^2)}, \quad 0 < q_\alpha < \frac{7}{5}, \tag{37}$$

and

$$\eta_{q,4} = \frac{2}{(7-5q_\alpha)} \frac{\omega_{Be} n_\alpha k_B T_\alpha}{(v_\alpha^2 + \omega_{Be}^2)}, \quad 0 < q_\alpha < \frac{7}{5}. \tag{38}$$

It can be proved that in the other calculations for the matrix elements in Eqs.(24), (25), (27) and (28), we can obtain the same results as Eqs.(32)-(35), (37) and (38).

In the above six viscosity coefficients, $\eta_{q,0}$ and $\zeta_q$ is not affected by the magnetic field, which are the same as the first and second viscosity coefficient, respectively, in [42]. $\eta_{q,i}$ ($i$=1,2,3,4) are the viscosity coefficients introduced by the magnetic field. As usual, $\eta_{q,3}$ and $\eta_{q,4}$ are also the gyroviscosity coefficients [46]. If the magnetic field was absence, $\eta_{q,3}$ and $\eta_{q,4}$ disappear, and in this case we have $\eta_{q,1}=\eta_{q,2}=\eta_{q,0}$, which so recovers to the previous case without magnetic field in [42].

Eqs.(32)-(35,) (37) and (38) show clearly the effects of magnetic field on the viscosity coefficients, which depend strongly on the $q$-parameter and magnetic field of the plasma with the power-law $q$-distributions. We can see that when we take $q_\alpha$=1, the viscosity coefficients all recover to the standard expresses for the plasma with a Maxwell distribution [45,46], respectively, the first viscosity coefficients,

$$\eta_{1,0} = \frac{n_\alpha k_B T_\alpha}{v_\alpha}, \quad \eta_{1,1} = \frac{v_\alpha n_\alpha k_B T_\alpha}{\left(v_\alpha^2 + 4\omega_{Be}^2\right)}, \quad \eta_{1,2} = \frac{v_\alpha n_\alpha k_B T_\alpha}{\left(v_\alpha^2 + \omega_{Be}^2\right)}, \tag{39}$$

the gyroviscosity coefficients,

$$\eta_{1,3} = \frac{2\omega_{Be} n_\alpha k_B T_\alpha}{\left(v_\alpha^2 + 4\omega_{Be}^2\right)}, \quad \eta_{1,4} = \frac{\omega_{Be} n_\alpha k_B T_\alpha}{\left(v_\alpha^2 + \omega_{Be}^2\right)}, \tag{40}$$

and the second viscosity coefficient,

$$\zeta_1 = \frac{5 n_\alpha k_B T_\alpha}{3 v_\alpha}. \tag{41}$$

## 4. Numerical analyses for the effects of magnetic field and nonextensivity

In order to show the roles of magnetic field in the viscosity of charged particles in the weakly ionized and magnetized plasma with power-law velocity $q$-distributions more clearly, we make numerical analyses for the viscosity coefficients $\eta_{q,i}$ ($i$=1,2,3,4) introduced by the magnetic field. For this purpose, from (33),(34), (37), (38) and (39), we write that

$$\frac{\eta_{q,1}}{\eta_{1,0}} = \frac{2}{(7-5q_\alpha)}\left[1+4\frac{\omega_{Be}^2}{v_\alpha^2}\right]^{-1}, \tag{40}$$

$$\frac{\eta_{q,2}}{\eta_{1,0}} = \frac{2}{(7-5q_\alpha)}\left[1+\frac{\omega_{Be}^2}{v_\alpha^2}\right]^{-1}, \tag{41}$$

$$\frac{\eta_{q,3}}{\eta_{1,0}} = \frac{4\omega_{Be}}{(7-5q_\alpha)v_\alpha}\left[1+4\left(\frac{\omega_{Be}^2}{v_\alpha^2}\right)\right]^{-1}, \tag{42}$$

and

$$\frac{\eta_{q,4}}{\eta_{1,0}} = \frac{2\omega_{Be}}{(7-5q_\alpha)v_\alpha}\left[1+\left(\frac{\omega_{Be}^2}{v_\alpha^2}\right)\right]^{-1}, \tag{43}$$

where denominator on the left-hand side of each equation in Eqs.(40)-(43) takes the first viscosity coefficient in the plasma with a Maxwell distribution and without magnetic field.

On the basis of the above equations Eqs.(40)-(43), in Figs.1-4, we give numerical analyses on the roles of magnetic field in the first viscosity coefficients, respectively, relative to that in the case of a Maxwell distribution and no magnetic field, and the calculation is for three different values of the $q$-parameters, $q_\alpha$=0.2, 1, and 1.2, where $\eta_{q,i}/\eta_{1,0}$, $i$=1,2,3,4, are respectively as the axis of ordinate and $\omega_{Be}/\nu_\alpha$ is the abscissa axis.

In Figs.1 and 2, the results show that the viscosity coefficients $\eta_{q,1}$ and $\eta_{q,2}$ will both decrease monotonously as the magnetic field increases. But in Figs.3 and 4, the results show that as the magnetic field increases, the gyroviscosity coefficients $\eta_{q,3}$ and $\eta_{q,4}$ both start to increase, reach a maximum, and then decrease. In all the four viscosity coefficients, we see that the $q$-parameter different from unity play a significant role. And they all will increase generally as the $q$-parameter increases.

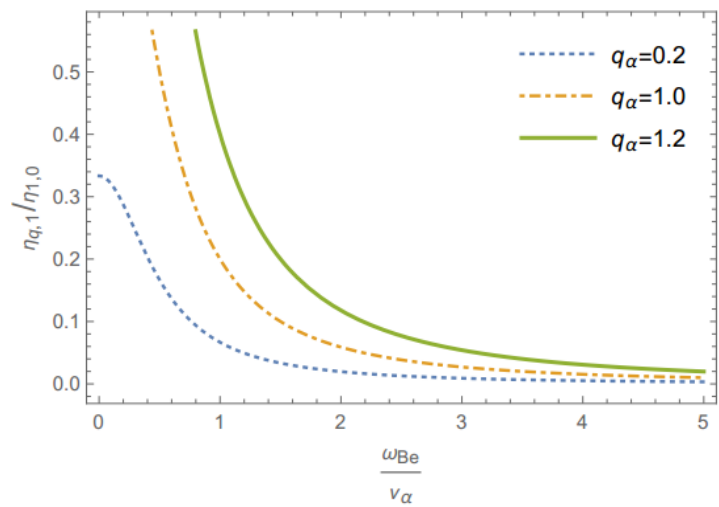

Fig.1 The role of magnetic field in $\eta_{q,1}$

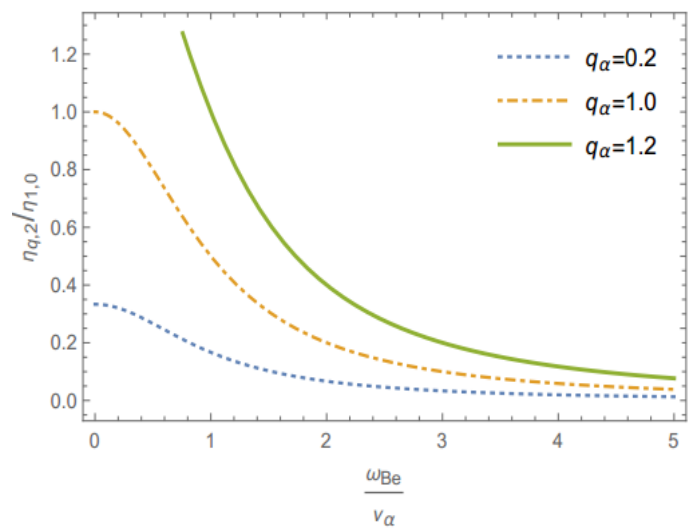

Fig.2 The role of magnetic field in $\eta_{q,2}$

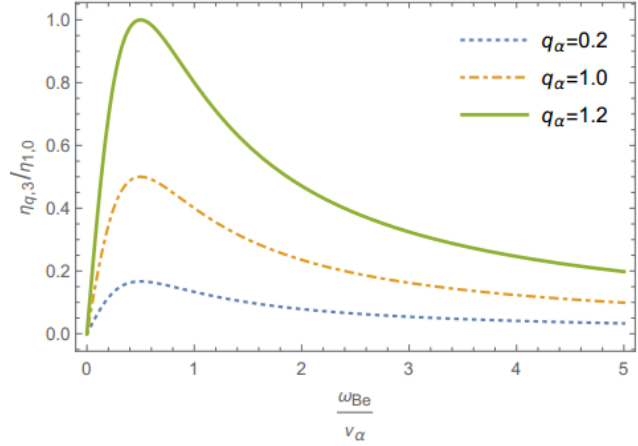

Fig.3 The role of magnetic field in $\eta_{q,3}$

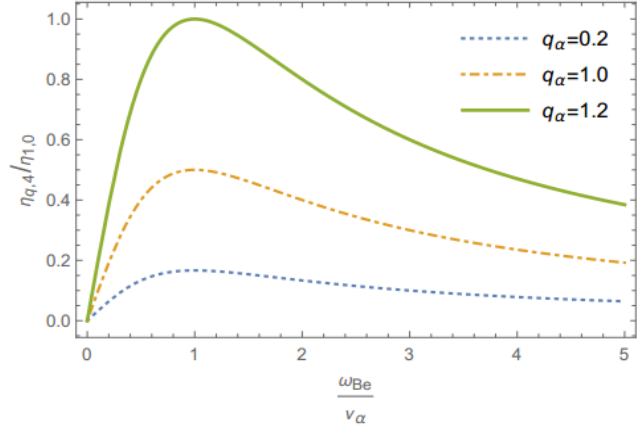

Fig.4 The role of magnetic field in $\eta_{q,4}$

**5. Conclusion**

In conclusion, we have studied the effect of magnetic field on the viscosity of charged particles in the weakly ionized and magnetized plasma with power-law $q$-distributions in nonextensive statistics. By using the generalized Boltzmann transport equations, we derived an expression (20) of the velocity $q$-distribution function of the plasma in the first-order approximation of Chapman-Enskog expansion. And then using the motion equations of hydrodynamics in the magnetized plasma, we investigated the matrix elements of the stress tensor in the plasma with power-law $q$-distributions and therefore we derived expressions of the viscosity coefficients of both electrons and ions, respectively including the first viscosity coefficients in Eqs.(32)-(34), the gyroviscosity coefficients in Eqs.(37) and (38), and the second viscosity coefficient in Eq.(35), where, as compared with the case without magnetic field, the magnetic field introduced four new viscosity coefficients $\eta_{q,1}$, $\eta_{q,2}$, $\eta_{q,3}$ and $\eta_{q,4}$, i.e., Eqs.(33), (34), (37) and (38). But the viscosity coefficient $\eta_{q,0}$ parallel to the magnetic field, Eq.(32), and the second viscosity coefficient $\zeta_q$, Eq.(35), are not affected by the magnetic field. If the magnetic field was removed, $\eta_{q,1}$ will equal to $\eta_{q,2}$ and they recover to $\eta_{q,0}$, and the gyroviscosity coefficients $\eta_{q,3}$ and $\eta_{q,4}$, Eqs.(37) and (38), will vanish.

All the six viscosity coefficients depend strongly on the nonextensive $q$-parameter and when we take the limit $q_\alpha \to 1$, they perfectly recover to those in the case of the plasma with a Maxwellian distribution.

By numerical analyses, the effects of magnetic field on the viscosity coefficients are shown clearly in Figs.1-4. And the roles of the $q$-parameter in the four coefficients are also be analyzed for three values of $q_\alpha = 0.2$, 1 and 1.2, respectively. We show that the viscosity coefficients $\eta_{q,1}$ and $\eta_{q,2}$ will both decrease monotonously as the magnetic field increases, but as the magnetic field increases, the gyroviscosity coefficients $\eta_{q,3}$ and $\eta_{q,4}$ both start to increase, reach a maximum, and then decrease. All the four viscosity coefficients will increase generally as the $q$-parameter increases.

Finally, in the coefficients Eqs.(32)-(35) (37) and (38), we have the limitation of value for the $q$-parameter, i.e., $0 < q_\alpha < 7/5$. In fact, such limitation for the $q$-parameter exists generally in the state equation in nonextensive statistics when the average in velocity space is made using the standard definition [47,48] or the $q$-average [49,50].

**Acknowledgement**

This work is supported by the National Natural Science Foundation of China under Grant No. 11775156.

**Appendix A**

Substituting Eq. (12) into Eq. (11), we obtain that

$$\left(v_\alpha k_{q1} - \omega_{Be} k_{q4}\right) v_x^2 + \left(v_\alpha k_{q2} + \omega_{Be} k_{q4}\right) v_y^2 + v_\alpha k_{q3} v_z^2 + \left[v_\alpha k_{q4} + 2\omega_{Be}\left(k_{q1} - k_{q2}\right)\right] v_x v_y$$
$$+ \left(v_\alpha k_5 - \omega_{Be} k_6\right) v_x v_z + \left(v_\alpha k_{q6} + \omega_{Be} k_{q5}\right) v_y v_z$$
$$= -\frac{m_\alpha}{k_B T_\alpha}\left[v_x^2 \frac{\partial u_x}{\partial x} + v_y^2 \frac{\partial u_y}{\partial y} + v_z^2 \frac{\partial u_z}{\partial z} + v_x v_y \left(\frac{\partial u_x}{\partial y} + \frac{\partial u_y}{\partial x}\right) + v_x v_z \left(\frac{\partial u_x}{\partial z} + \frac{\partial u_z}{\partial x}\right)\right.$$
$$\left. + v_y v_z \left(\frac{\partial u_y}{\partial z} + \frac{\partial u_z}{\partial y}\right)\right]\left[1 - (1-q_\alpha)\frac{m_\alpha \mathbf{v}^2}{2 k_B T_\alpha}\right]^{-1} f_{q,\alpha}. \qquad (A.1)$$

Comparing two sides of Eq.(A.1), we find the following equations,

$$v_\alpha k_{q1} - \omega_{Be} k_{q4} = -\frac{m_\alpha f_{q,\alpha}}{k_B T_\alpha}\left[1 - (1-q_\alpha)\frac{m_\alpha v^2}{2 k_B T_\alpha}\right]^{-1} \frac{\partial u_x}{\partial x},$$

$$v_\alpha k_{q2} + \omega_{Be} k_{q4} = -\frac{m_\alpha f_{q,\alpha}}{k_B T_\alpha}\left[1 - (1-q_\alpha)\frac{m_\alpha v^2}{2 k_B T_\alpha}\right]^{-1} \frac{\partial u_y}{\partial y},$$

$$v_\alpha k_{q3} = -\frac{m_\alpha f_{q,\alpha}}{k_B T_\alpha}\left[1 - (1-q_\alpha)\frac{m_\alpha v^2}{2 k_B T_\alpha}\right]^{-1} \frac{\partial u_z}{\partial z},$$

$$v_\alpha k_{q4} + 2\omega_{Be}\left(k_{q1} - k_{q2}\right) = -\frac{m_\alpha f_{q,\alpha}}{k_B T_\alpha}\left[1 - (1-q_\alpha)\frac{m_\alpha v^2}{2 k_B T_\alpha}\right]^{-1}\left(\frac{\partial u_x}{\partial y} + \frac{\partial u_y}{\partial x}\right),$$

$$v_\alpha k_{q5} - \omega_{Be} k_{q6} = -\frac{m_\alpha f_{q,\alpha}}{k_B T_\alpha}\left[1-(1-q_\alpha)\frac{m_\alpha v^2}{2k_B T_\alpha}\right]^{-1}\left(\frac{\partial u_x}{\partial z}+\frac{\partial u_z}{\partial x}\right),$$

and

$$v_\alpha k_{q6} + \omega_{Be} k_{q5} = -\frac{m_\alpha f_{q,\alpha}}{k_B T_\alpha}\left[1-(1-q_\alpha)\frac{m_\alpha v^2}{2k_B T_\alpha}\right]^{-1}\left(\frac{\partial u_y}{\partial z}+\frac{\partial u_z}{\partial y}\right). \tag{A.2}$$

Solving these equations in (A.2), we can easily determine these functions given by Eqs.(13)-(18).

**Appendix B**

Eq. (30) is calculated as follows.

$$\begin{aligned}P_{\alpha,xx} &= m_\alpha \int d\mathbf{v}\, v_x^2 \left(f_{q,\alpha} + k_{q1}v_x^2 + k_{q2}v_y^2 + k_{q3}v_z^2 + k_{q4}v_x v_y + k_{q5}v_x v_z + k_{q6}v_y v_z\right) \\ &= m_\alpha \int d\mathbf{v}\, v_x^2 \left(f_{q,\alpha} + k_{q1}v_x^2 + k_{q2}v_y^2 + k_{q3}v_z^2\right) \\ &= m_\alpha \int v_x^2 f_{q,\alpha} d\mathbf{v} + \int \frac{m_\alpha v_x^2}{v_\alpha} F_{q,\alpha}(v) V_\alpha(v) d\mathbf{v},\end{aligned} \tag{B.1}$$

where

$$F_{q,\alpha}(v) = -\frac{m_\alpha f_{q,\alpha}}{k_B T_\alpha \left(v_\alpha^2 + 4\omega_{Be}^2\right)}\left[1-(1-q_\alpha)\frac{m_\alpha v^2}{2k_B T_\alpha}\right]^{-1},$$

$$V_\alpha(v) = \left[\left(v_\alpha^2 + 2\omega_{Be}^2\right)\frac{\partial u_x}{\partial x} + 2\omega_{Be}^2\frac{\partial u_y}{\partial y} + \omega_{Be}v_\alpha\left(\frac{\partial u_x}{\partial y}+\frac{\partial u_y}{\partial x}\right)\right]v_x^2$$
$$+\left[2\omega_{Be}^2\frac{\partial u_x}{\partial x} + \left(v_\alpha^2 + 2\omega_{Be}^2\right)\frac{\partial u_y}{\partial y} - \omega_{Be}v_\alpha\left(\frac{\partial u_x}{\partial y}+\frac{\partial u_y}{\partial x}\right)\right]v_y^2 + \left(v_\alpha^2 + 4\omega_{Be}^2\right)\left(\frac{\partial u_z}{\partial z}\right)v_z^2.$$

Furthermore, we derive that

$$\begin{aligned}P_{\alpha,xx} &= \frac{m_\alpha}{3}\int v^2 f_{q,\alpha} d\mathbf{v} + \int \frac{m_\alpha C_\alpha}{15 v_\alpha} v^4 F_{q,\alpha}(v) d\mathbf{v} = \frac{4\pi m_\alpha}{3}\int v^4 f_{q,\alpha} dv + \int \frac{4\pi m_\alpha C_\alpha}{15 v_\alpha} v^6 F_{q,\alpha}(v) dv, \\ &= \frac{4\pi}{3} m_\alpha n_\alpha B_{q,\alpha}\left(\frac{m_\alpha}{2\pi k_B T_\alpha}\right)^{\frac{3}{2}}\int v^4\left[1-(1-q_\alpha)\frac{m_\alpha v^2}{2k_B T_\alpha}\right]^{\frac{1}{1-q_\alpha}} dv \\ &\quad - \frac{4\pi m_\alpha^2 n_\alpha B_{q,\alpha} C_\alpha}{15 v_\alpha k_B T_\alpha \left(v_\alpha^2 + 4\omega_{Be}^2\right)}\left(\frac{m_\alpha}{2\pi k_B T_\alpha}\right)^{\frac{3}{2}}\int v^6\left[1-(1-q_\alpha)\frac{m_\alpha v^2}{2k_B T_\alpha}\right]^{\frac{1}{1-q_\alpha}-1} dv.\end{aligned} \tag{B.2}$$

where

$$C_\alpha = \left(3v_\alpha^2 + 8\omega_{Be}^2\right)\frac{\partial u_x}{\partial x} + \left(v_\alpha^2 + 8\omega_{Be}^2\right)\frac{\partial u_y}{\partial y} + \left(v_\alpha^2 + 4\omega_{Be}^2\right)\frac{\partial u_z}{\partial z} + 2\omega_{Be}v_\alpha\left(\frac{\partial u_x}{\partial y}+\frac{\partial u_y}{\partial x}\right).$$

For $q_\alpha > 1$, the integrals in Eq.(B.2) are calculated as

$$\int_0^{+\infty} v^4\left[1-(1-q_\alpha)\frac{m_\alpha v^2}{2k_B T_\alpha}\right]^{\frac{1}{1-q_\alpha}} dv = 3\sqrt{\frac{\pi}{2}}\left[\frac{m_\alpha(q_\alpha-1)}{k_B T_\alpha}\right]^{-\frac{5}{2}}\frac{\Gamma\left(-\frac{5}{2}+\frac{1}{q_\alpha-1}\right)}{\Gamma\left(\frac{1}{q_\alpha-1}\right)}, \quad 1 < q_\alpha < \frac{7}{5}, \tag{B.3}$$

$$\int_0^{+\infty} v^6 \left[1-(1-q_\alpha)\frac{m_\alpha v^2}{2k_B T_\alpha}\right]^{\frac{1}{1-q_\alpha}-1} dv = 15\sqrt{\frac{\pi}{2}} \left[\frac{m_\alpha(q_\alpha-1)}{k_B T_\alpha}\right]^{\frac{7}{2}} \frac{\Gamma\left(-\frac{5}{2}+\frac{1}{q_\alpha-1}\right)}{\Gamma\left(\frac{q_\alpha}{q_\alpha-1}\right)}, \quad 1<q_\alpha<\frac{7}{5}. \quad (B.4)$$

And therefore we obtain that

$$P_{\alpha,xx} = \frac{2n_\alpha k_B T_\alpha}{(7-5q_\alpha)} - \frac{2n_\alpha k_B T_\alpha C_\alpha}{(7-5q_\alpha)(v_\alpha^2 + 4\omega_{Be}^2)v_\alpha}. \quad (B.5)$$

For $0 < q_\alpha < 1$, there is a thermal cutoff. If $a = \sqrt{\frac{2k_B T_\alpha}{m_\alpha(1-q_\alpha)}}$, the integrals in (B.2) are calculated as

$$\int_0^a v^4 \left[1-(1-q_\alpha)\frac{m_\alpha v^2}{2k_B T_\alpha}\right]^{\frac{1}{1-q_\alpha}} dv = 3\sqrt{\frac{\pi}{2}} \left[\frac{k_B T_\alpha}{m_\alpha(1-q_\alpha)}\right]^{\frac{5}{2}} \frac{\Gamma\left(1+\frac{1}{1-q_\alpha}\right)}{\Gamma\left(\frac{7}{2}+\frac{1}{1-q_\alpha}\right)}, \quad (B.6)$$

$$\int_0^a v^6 \left[1-(1-q_\alpha)\frac{m_\alpha v^2}{2k_B T_\alpha}\right]^{\frac{1}{1-q_\alpha}-1} dv = 15\sqrt{\frac{\pi}{2}} \left[\frac{k_B T_\alpha}{m_\alpha(1-q_\alpha)}\right]^{\frac{7}{2}} \frac{\Gamma\left(\frac{1}{1-q_\alpha}\right)}{\Gamma\left(\frac{7}{2}+\frac{1}{1-q_\alpha}\right)}. \quad (B.7)$$

And therefore we obtain that

$$P_{\alpha,xx} = \frac{2n_\alpha k_B T_\alpha}{(7-5q_\alpha)} - \frac{2n_\alpha k_B T_\alpha C_\alpha}{(7-5q_\alpha)(v_\alpha^2 + 4\omega_{Be}^2)v_\alpha}. \quad (B.8)$$